\documentclass[preprint,showpacs,preprintnumbers,amsmath,amssymb]{revtex4}
\usepackage{amssymb}

\usepackage{graphicx}
\usepackage{dcolumn}
\usepackage{bm}

\begin{document}

\title{Experimental violation of the Leggett-Garg inequality under decoherence}

\author{Jin-Shi Xu, Chuan-Feng Li$\footnote{
email: cfli@ustc.edu.cn}$, Xu-Bo Zou, and Guang-Can Guo}
 \affiliation{Key
Laboratory of Quantum Information, University of Science and
Technology of China, CAS, Hefei, 230026, People's Republic of China}

\date{\today}

\begin{abstract}
Despite the great success of quantum mechanics, questions regarding its application still exist and the boundary between quantum and classical mechanics remains unclear. Based on the philosophical assumptions of macrorealism and noninvasive measurability, Leggett and Garg devised a series of inequalities (LG inequalities) involving a single system with a set of measurements at different times. Introduced as the Bell inequalities in time, the violation of LG inequalities excludes the hidden-variable description based on the above two assumptions. We experimentally investigated the single photon LG inequalities under decoherence simulated by birefringent media. These generalized LG inequalities test the evolution trajectory of the photon and are shown to be maximally violated in a coherent evolution process. The violation of LG inequalities becomes weaker with the increase of interaction time in the environment. The ability to violate the LG inequalities can be used to set a boundary of the classical realistic description.
\end{abstract}

\maketitle

\section{Introduction}
The theory of quantum mechanics has proven to be very successful. The theory not only provides precise explanations of many physical phenomena, but also has resulted in the development of many modern technologies \cite{Zeilinger00}. However, questions regarding the applicability of quantum mechanics to macroscopic systems still exist, and the boundary between quantum and classical mechanics remains unclear. The association between classical mechanics and macroscopic systems was tentatively accepted during the early development of quantum mechanics theory \cite{Zurek03}. This viewpoint is embodied in a famous paradox proposed by Schr\"{o}dinger in 1935
\cite{Schrodinger35}, in which he described a ``quite absurd" example that a cat state may be alive and dead at the same time. Nowadays, the difficulty of observing the Schr\"{o}dinger cat state is explained by decoherence, where the superposition of distinct states is destroyed by coupling with unwanted degrees of freedom \cite{Zurek032}.

Leggett-Garg inequalities (LG inequalities) have been derived to clarify the validity of generalizing quantum mechanics to macroscopic systems, based on the macrorealistic theory with macrorealism and noninvasive measurability assumptions \cite{Leggett85}. These inequalities involve a single system with a set of measurements at different times and play a role similar to that of the Bell inequalities in testing local hidden-variable theories \cite{Bell64}. Introduced as the Bell inequalities in time, the violation of LG inequalities excludes the hidden-variable description based on the above two assumptions.

The two assumptions of the LG inequalities can be extended to any physical system under the classical realistic description if the philosophy of macrorealism is divorced from macroscopic objects. In such descriptions, the state of a system with two or more distinct states will at all times be in one or the other of these states (macrorealism). A corollary of this is that it is possible to determine the state of a system without any disturbance of its subsequent dynamics (noninvasive measurability). The original proposal to realise noninvasive measurement, by coupling the interested system to a probe \cite{Leggett85}, is similar to the Controlled-Not (CNOT) gate where an ancilla is used as the target qubit and the interested system as the control qubit \cite{Nielsen00}.

In this study, we experimentally investigate the single photon LG inequalities in a dephasing environment simulated by birefringent media. By implementing an optical CNOT gate on a single photon, the LG inequalities are shown to be maximally violated in a coherent evolution process. This disproves its classical realistic description with the two assumptions of the LG inequalities. With the increase of birefringent media, the violation of LG inequalities becomes weaker and is shown to be not violated anymore at some time. The ability to violate the LG inequalities can be used to set the boundary of the classical realistic description.

\section{Results}
\begin{figure}[tbph]
\begin{center}
\includegraphics[width= 3.3in]{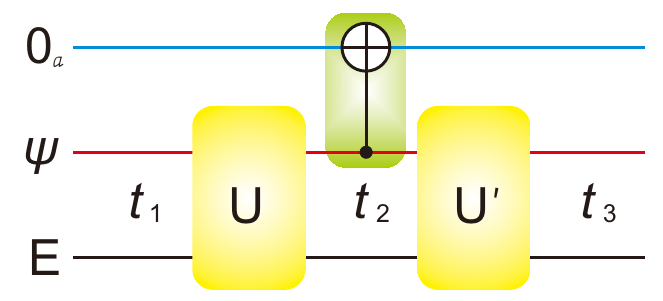}
\end{center}
\caption{Logic circuit to measure the value of $K(t_{2},t_{3})$ with
a CNOT gate. $0_{a}$ is the initial state of the ancilla.
$\psi$ is the state of the system (can only be in 0 or 1 during the evolution under the classical realistic description). E represents the
environment with operation of U between $t_{1}$ and $t_{2}$, and
U$'$ between $t_{2}$ and $t_{3}$, respectively.} \label{logic}
\end{figure}

{\bf Theoretical Schemes.} Consider an observable $Q(t)$ of a two level physical system, where
$ \left\vert 0 \right\rangle $ and $\left\vert 1 \right\rangle$ are
the two eigenstates of $Q(t)$ with the eigenvalues of +1 and -1, respectively. The two
different times correlation function of this observable is defined
as $K(t_{1},t_{2})=\langle Q(t_{1})Q(t_{2})\rangle$. For
three different times $t_{1}$, $t_{2}$ and $t_{3}$ (using the same
deduction of Huelga \textit{et al.} \cite{Huelga95}), we can obtain the following:
\begin{equation}
 K(t_{1},t_{3})- K(t_{1},t_{2})- K(t_{2},t_{3})\geq-1 \text{,}
 \label{2}
\end{equation}
\begin{equation}
K(t_{1},t_{3})+ K(t_{1},t_{2})+ K(t_{2},t_{3})\geq-1 \text{.}
  \label{3}
\end{equation}
These two inequalities are the Wigner type L-G inequalities
\cite{Wigner70,Kofler08} under the classical realistic description with the two assumptions. To experimentally verify them, the values of $K(t_{1},t_{2})$, $K(t_{2},t_{3})$ and
$K(t_{1},t_{3})$ should be measured. If we choose $t_{1}$ as the
initial time, i.e. $t_{1}=0$, we can conveniently used projective
measurement at $t_{2}$ or $t_{3}$ to get $K(t_{1},t_{2})$ and
$K(t_{1},t_{3})$, because the dynamics after $t_{2}$ or
$t_{3}$ are not of interest in these two cases. While measuring
$K(t_{2},t_{3})$, we implemented a CNOT operation that has the ability to realize noninvasive measurement under the classical realistic description at $t_{2}$
and projective measurement at $t_{3}$. Figure \ref{logic} shows the logic circuit. The two-level ancillary state was initially prepared into the ground state $0_{a}$. The system of interest with initial state $\psi$ (either $0$ or $1$) evolves in the environment E with an operation of U between $t_{1}$ and $t_{2}$ and U' between $t_{2}$ and $t_{3}$. At time $t_{2}$, the physical control system was coupled to the ancilla, which was used as the target system. If the state of $\psi$ is 0, the ancilla remains in $0_{a}$ without any change. Otherwise, the state of the ancilla system will be flipped and changed to the excited state $1_{a}$. As a result, by detecting the state of the ancilla, we can know the state of $\psi$ at time $t_{2}$.

{\bf Experimental violation of the generalized Leggett-Garg inequalities under decoherence.} Photon qubits which is easily manipulated at the single qubit level
and can be excellently isolated from the environment, play important
roles in quantum communication and quantum computation
\cite{Gisin02,Kok07}. The optical CNOT gate has been used to make a strong coupling to an ancilla, for the purpose of measurement of a signal \cite{Huang01,Barbieri07,Sciarrino06,Pryde04}. By encoding a single photon with several qubits, the CNOT gate can be readily realized with simple optical components \cite{Cerf98}. Moreover, by introducing birefringent quartz plates where the coupling between the photon's polarization and frequency modes occurs, we can simulate a fully controllable ``environment" to
investigate the evolution of the photon state \cite{Berglund00}.
Here, we encode the observable $Q(t)$ as the polarization of a single
photon, where the $45^{\circ}$ linear polarization state
$|\overline{H}\rangle=\frac{1}{\sqrt{2}}(|H\rangle+|V\rangle)$ (
$|H\rangle$ and $|V\rangle$ represent the horizontal and vertical
polarization states respectively) is used as $|0\rangle $ with the
eigenvalue of +1 and the $-45^{\circ}$ linear polarization state
$|\overline{V}\rangle=\frac{1}{\sqrt{2}}(|H\rangle-|V\rangle)$ as
$|1\rangle$ with the eigenvalue of -1. As a result, the observable of $Q(t)$ is the Pauli $\sigma_{x}$ operator. In our experiment, we use the
herald single photon source produced from the pulsed parametric
down-conversion process in a nonlinear crystal \cite{Pittman05}. In
this process, one of the photon is used as the trigger, while the
other is prepared to be $|\overline{H}\rangle$ and used as the
initial input state (see Methods for details).

\begin{figure}[tbph]
\begin{center}
\includegraphics[width= 3.3in]{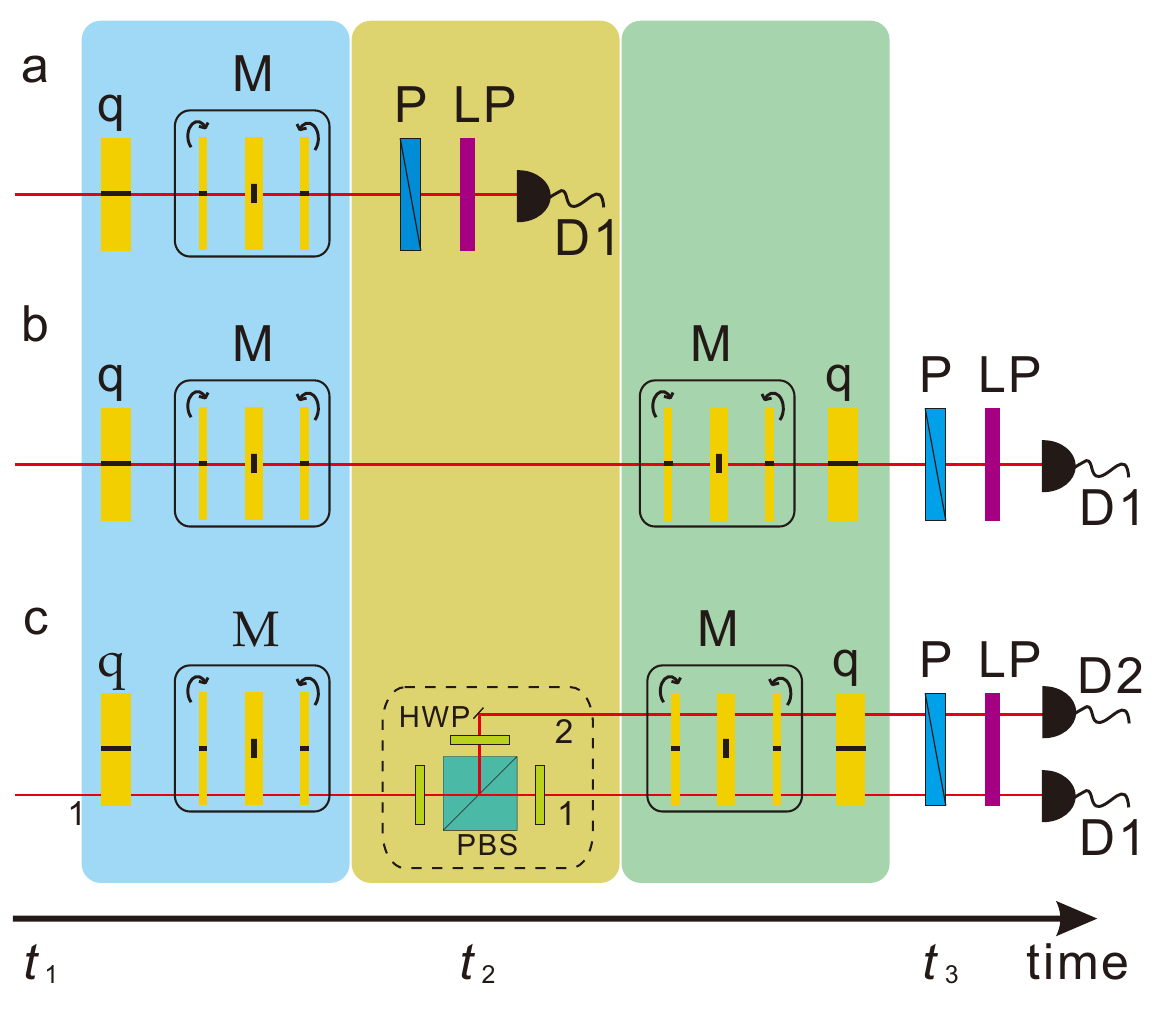}
\end{center}
\caption{(Color on line). (a) The setup to measure $K(t_{1},t_{2})$. The quartz plate (q) with the tiltable combination of quartz plates in the solid pane M represents the evolution environment (black bars represent the optic axes of the quartz). The final measurement basis was chosen by the polarizer (P). The photon was coupled by a multimode fibre to the single photon detector D1 equipped with a long pass lens (LP) in front of it. (b) The setup to measure $K(t_{2},t_{3})$. Two equal settings of quartz plates of q and M are used to simulate the environment operators U and U' in fig. 1. (c) The setup to measure $K(t_{2},t_{3})$. The dashed pane containing a polarization beam splitter (PBS) and three half wave plates (HWP) with optic axes set at $22.5^{\circ}$ was inserted at time $t_{2}$. The single photon detector D2 is used to detect the photon in path 2.}
\label{setup}
\end{figure}

Fig. \ref{setup} shows the experimental setup for investigating the
evolution of the interested photon. Figure \ref{setup}a shows the setup to measure the value of $K(t_{1},t_{2})$. The quartz plate q and a tiltable combination of quartz plates M represent the evolution environment (the total thickness of quartz plates is $L$). The solid pane M contains two parallel quartz plates (optic axes are set to be horizontal) with the thickness of 8$\lambda_{0}$ (0.78 $\mu$m) and a mutual perpendicular quartz plate with the thickness of 16$\lambda_{0}$, where black bars represent the direction of their optic axes. By tilting these two 8$\lambda_{0}$ quartz plates, we can introduce the required relative phase between $H$ and $V$. The measurement basis is chosen by a polarizer (P). The photon is then coupled by a multimode fibre to the single photon detector D1 equipped with a Long pass lenses (LP) in front of it, which is used to minimise the background caused by the pump beam light. Figure 2b represents the setup to measure $K(t_{1},t_{3})$ with two equal sets of quartz plates of q and M, in which the evolution time is twice of that in Fig. 2a. In our setting, the evolution from $t_{1}$ to $t_{2}$ is the same as that from $t_{2}$ to $t_{3}$ (the time duration is denoted as $t$), which means U=U'. In order to measure $K(t_{2},t_{3})$, the dashed pane, containing a polarization beam splitter (PBS) and three half wave plates (HWP), with optic axes set at $22.5^{\circ}$, is implemented at time $t_{2}$ as shown in Fig. 2c. The dashed pane transmits the $45^{\circ}$ polarization state (path 1) and reflects the $-45^{\circ}$ polarization state (path 2). As a result, if the ancilla qubit is encoded as the path information of the photon, the dashed pane acts as the CNOT gate with the path of the photon used as the target qubit and the polarization as the control qubit. Another single photon detector (D2) is applied to detect the photon in the path 2.

We first analysed the single photon LG inequalities under the
classical realistic description with the two assumptions, where
the system can only be in one of these two states $\overline{H}$ and
$\overline{V}$. If the input photon is initially in the state
$\rho_{0}=\overline{H}$, after evolution time $t$, the state becomes
$\rho_{t}=(1-\alpha)\overline{H}+\alpha\overline{V}$, where $\alpha$
represents the influence of the environment (i.e., the probability
of the photon flips from $\overline{H}$ to $\overline{V}$ and it is a
function of $t$ with $0\leq\alpha\leq1$). With further identical
interaction time $t$ in the same environment, the final state
evolves to
$\rho_{2t}=(\alpha^{2}+(1-\alpha)^{2})\overline{H}+2\alpha(1-\alpha)\overline{V}$.
Therefore,
$K(t_{1},t_{2})=P_{\overline{H}_{1},\overline{H}_{2}}-P_{\overline{H}_{1},\overline{V}_{2}}=1-2\alpha$
and
$K(t_{1},t_{3})=P_{\overline{H}_{1},\overline{H}_{3}}-P_{\overline{H}_{1},\overline{V}_{3}}=4\alpha^{2}-4\alpha+1$,
where $P_{G_{i},O_{j}}$ ($G, O\in\{\overline{H},\overline{V}\}, i,
j\in\{1,2,3\}$) represent the probability of detecting $O$ at time
$t_{j}$ when the initial state is $G$ at time $t_{i}$. While for
$K(t_{2},t_{3})$, with the CNOT operation at time $t_{2}$, we have
the probability of $1-\alpha$ to get $\overline{H}$. After another
evolution time $t$, the final state is the same as $\rho_{t}$. We
also have the probability of $\alpha$ to get $\overline{V}$ and the
subsequent state becomes
$\rho_{t}^{'}=(1-\alpha)\overline{V}+\alpha\overline{H}$. As a
result, we can get
$K(t_{2},t_{3})=P_{\overline{H}_{2}}(P_{\overline{H}_{2},\overline{H}_{3}}-P_{\overline{H}_{2},\overline{V}_{3}})
+P_{\overline{V}_{2}}(P_{\overline{V}_{2},\overline{V}_{3}}-P_{\overline{V}_{2},\overline{H}_{3}})=1-2\alpha$,
where $P_{G_{i}}$ represents the probability to detect $G$ at time
$t_{i}$. It is then easy to verify that
$K(t_{1},t_{3})-(K(t_{1},t_{2})+K(t_{2},t_{3}))+1=4\alpha^{2}\geq0$
and
$K(t_{1},t_{3})+(K(t_{1},t_{2})+K(t_{2},t_{3}))+1=4(\alpha-1)^{2}\geq0$
for any $\alpha$. Therefore, the inequalities (1) and (2) are
trivial results in the classical realistic description.

Next, we analysed the experiment from the viewpoint of quantum
mechanics. For the case of coherence evolution, the evolution effect
is imposed by tilting the quartz in M. Because U=U$'$, the induced
relative phase is $\delta$ from evolution time $t_{1}$ to $t_{2}$ as
well as from $t_{2}$ to $t_{3}$ and the induced phase from $t_{1}
$to $t_{3}$ is $2\delta$. If the input state is
$|\overline{H}\rangle$, after passing the first solid pane M the
state becomes $\left\vert \psi _{t_{2}}\right\rangle
=\frac{1}{2}(1+e^{i\delta})\left\vert \ \overline{H}\right\rangle
+\frac{1}{2}(1-e^{i\delta})\left\vert \ \overline{V}\right\rangle$
and $K(t_{1},t_{2})=\cos\delta$. With the same analysis, we can get
$K(t_{1},t_{3})=\cos2\delta$. When measuring $K(t_{2},t_{3})$, if
the state is $|\overline{H}\rangle$ at time $t_{2}$, its subsequent
evolution state is the same as $\left\vert \psi
_{t_{2}}\right\rangle$; if the state is $|\overline{V}\rangle$, the
subsequent state becomes $\left\vert \psi _{t_{2}}^{'}\right\rangle
=\frac{1}{2}(1+e^{i\delta})\left\vert \ \overline{V}\right\rangle
+\frac{1}{2}(1-e^{i\delta})\left\vert \ \overline{H}\right\rangle$.
Therefore $K(t_{2},t_{3})=\cos\delta$. These two LG inequalities can
then be calculated as $K_{-}=\cos(2\delta)-2\cos(\delta)$ and
$K_{+}=\cos(2\delta)+2\cos(\delta)$. It can be seen that $K_{-}$
reaches its minimum $-1.5$ with $\delta=\frac{\pi}{3}$ and $K_{+}$
also reaches its minimum $-1.5$ with $\delta=\frac{2\pi}{3}$.

\begin{figure}[tbph]
\begin{center}
\includegraphics[width= 3in]{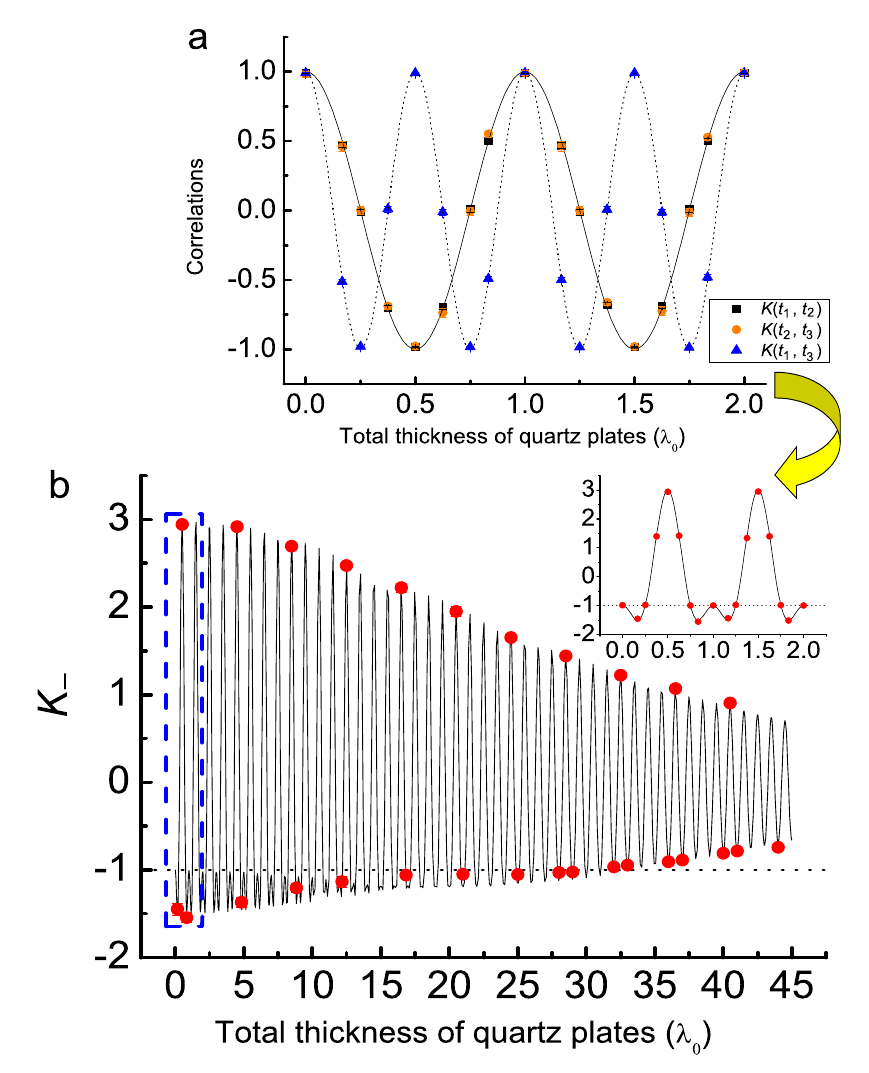}
\end{center}
\caption{(Color online). (a) The corresponding values for individual correlations $K(t_{1},t_{2})$, $K(t_{2},t_{3})$ and $K(t_{1},t_{3})$ to get $K_{-}$ in the inset of (b). The solid line, dashed line and dotted line are the corresponding theoretical predictions (the solid line and the dashed line completely overlap and only the solid line can be seen). The $x$ axis represents the total thickness of quartz plates between $t_{1}$ and $t_{2}$. (b) The envelope evolution of $K_{-}$. Red dots represent the experimental results. Solid lines are the theoretical fittings employing equation (3). The dashed line represents the classical limit, -1. The $x$ axis represents the total thickness of quartz plates between $t_{1}$ and $t_{2}$. The inset displays the oscillation between the maximum and minimum in the blue dashed pane (the $x$ axis represents the total thickness of quartz plates between $t_{1}$ and $t_{2}$, and the $y$ axis represents $K_{-}$). Error bars correspond to the random fluctuations of each measured coincidence count and the tilt uncertainties of quartz plates. $\lambda_{0}=0.78$ $\mu$m.} \label{fig:decoherence1}
\end{figure}

We further consider the decoherence evolution case, which is
achieved by increasing the thickness of quartz plates. In this case,
the frequency spectrum of the photon is considered as a Gaussian
amplitude function $f(\omega)$ with the central frequency
$\omega_{0}$ corresponding to the central wavelength 0.78 $\mu$m and
frequency spread $\sigma$. For a special frequency $\omega$, after the photon passes through the quartz
plates with thickness $L$, the
induced relative phase is $\gamma\omega$, where
$\gamma=L\Delta n/c$. $c$ represents the velocity of the photon in
the vacuum and $\Delta n$ is the difference between the indices of
refraction of ordinary and extraordinary light. With a trace over all the frequency modes, the final forms of LG
inequalities can be written as
\begin{equation}
K_{-}=\cos(2\gamma\omega_{\circ})\exp(-\frac{1}{4}\gamma^{2}\sigma^{2})-
2\cos(\gamma\omega_{\circ})\exp(-\frac{1}{16}\gamma^2\sigma^{2})\text{,}
\label{eq:decoherence1}
\end{equation}
\begin{equation}
K_{+}=\cos(2\gamma\omega_{\circ})\exp(-\frac{1}{4}\gamma^{2}\sigma^{2})+
2\cos(\gamma\omega_{\circ})\exp(-\frac{1}{16}\gamma^2\sigma^{2})\text{.}
\label{eq:decoherence2}
\end{equation}
Obviously, when the thickness $L$ is small, equations (\ref{eq:decoherence1}) and
(\ref{eq:decoherence2}) trend toward coherent evolution.

\begin{figure}[tbph]
\begin{center}
\includegraphics[width= 3in]{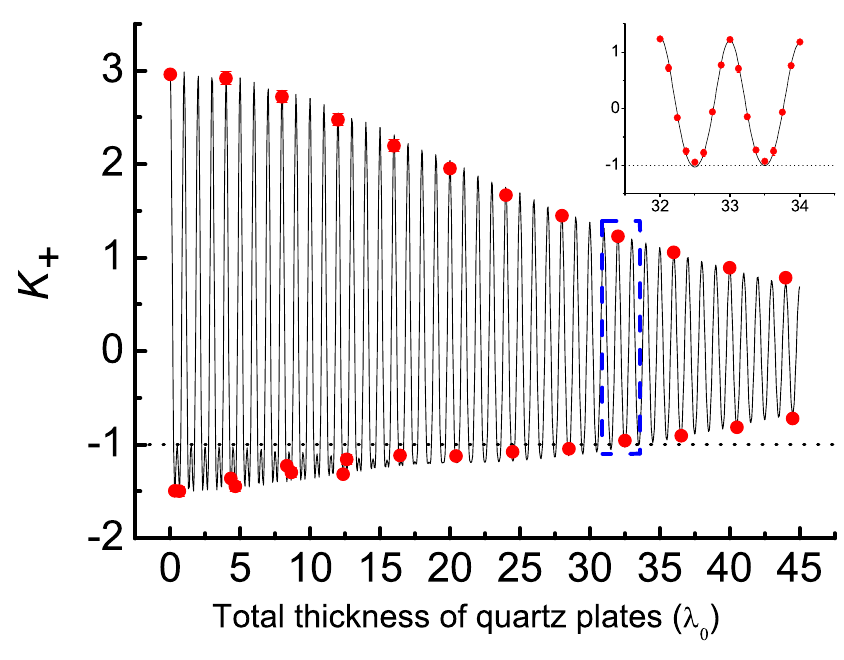}
\end{center}
\caption{(Color online). Red dots represent the experimental results. Solid lines are the theoretical fittings employing equation (4). The dashed line represents the classical limit, -1. The $x$ axis represents the total thickness of quartz plates between $t_{1}$ and $t_{2}$. The inset represents the oscillation between the maximum and minimum in the blue dashed pane (the $x$ axis represents the total thickness of quartz plates between $t_{1}$ and $t_{2}$, and the $y$ axis represents $K_{+}$). Error bars correspond to the random fluctuation of each measured coincidence count and the tilt uncertainties of quartz plates.}
\label{fig:decoherence2}
\end{figure}

Fig. 3a shows the corresponding values for individual correlations $K(t_{1},t_{2})$, $K(t_{2},t_{3})$ and $K(t_{1},t_{3})$, which are used to get the values of $K_{-}$ in the inset of Fig. 3b. The solid line, dashed line and dotted line correspond to theoretical predictions (the solid line and the dashed line completely overlap and only the solid line can be seen). We find that $K(t_{1},t_{2})=K(t_{2},t_{3})$ and the oscillation period of $K(t_{1},t_{3})$ is twice as that of $K(t_{1},t_{2})$ ($K(t_{2},t_{3})$). These findings are consistent with theoretical predictions. Fig. 3b shows the envelope evolution of $K_{-}$. When the thickness of quartz plates is small, the generalized LG inequality is violated according to the previous analysis. From the inset in Fig. 3b, which represents the oscillation between the maximum and minimum in the blue dashed pane, we find that the minimum of $K_{-}$ reaches $-1.544\pm0.056$ , which violates the classical limit of -1 by about 9.7 standard deviations. With the increase in thickness of quartz plates ($L$), the violation of the LG inequalities becomes gradually weaker. $K_{-}$ does not violate the classical limit -1 when $L$ is increased to about 33$\lambda_{0}$. This implies that when $L$ is larger than 33$\lambda_{0}$, the evolution trajectory can be described by the classical realistic description, and when $L$ is smaller than 33$\lambda_{0}$, the trajectory must adopt the quantum description. Therefore, we have set a boundary for the classical realistic description by using the LG inequalities. Errors are mainly due to the random fluctuation of each measured coincidence count and the tilt uncertainties of quartz plates (we tilt quartz plates to introduce the required relative phase between horizontal and vertical polarizations). Solid lines are the theoretical predictions of $K_{-}$, employing equation (3) with $\sigma$ fitting to $3.56\times10^{13}$ Hz.

We further show the envelope evolution of $K_{+}$ in Fig.
\ref{fig:decoherence2}. At the beginning of the evolution, the
minimal value of $K_{+}$ reaches $-1.495\pm0.052$ which violates the
LG inequality by about 10 standard deviations. When $L$ increases
to about $33\lambda_{0}$, it does not violate the classical limit
anymore. The inset shows the oscillation between the maximum
and minimum in the dashed pane, which displays the critical
boundary. Solid lines are the theoretical
predictions employing equation. (\ref{eq:decoherence2}).

\section{Discussion}
In our experiment, the polarization of a photon was used as the observable $Q(t)$. This measured quantity could also be considered as the evolution path of the photon. A photon with different polarizations passes through different paths, separated by the polarization beam splitter. This phenomenon is similar to that of the position of a single electron in a double quantum dot \cite{Jordan06}. The violation of generalized LG inequalities implies that at least one of the two assumptions in the classical realistic description is untenable and disproves the definite classical evolution trajectory \cite{Jordan06}. In our experiment, the information carrier (polarization) and the environment freedom (frequency) are encoded on the same photon. The experimental results can be repeated by a corresponding diagonally polarized input laser pulse, in which each of the photons in the laser pulse undergoes the same evolution. The polarization of a ``classical" light (laser pulse) can also be viewed as a consequence of the transverse vector of electromagnetic field that is allowed to be superposed, in which $Q(t)$ ranges continuously from -1 to 1. This condition is different from the initial assumption that $Q(t)$ can only be of 1 or -1 at each measurement in our case. As a result, the violation of LG inequalities with ``classical" light does not contradict that case of single photon.

Recently, the violation of generalized LG inequalities has been demonstrated by employing weak measurements on a single photon \cite{Goggin11} and a superconducting quantum circuit \cite{Palacios10}. The generalized LG inequalities used in these studies are similar to the Wigner version used here, which are derived from the classical realistic description with the two assumptions. The weak measurement provides the ability to control the back action on the system in the sense of quantum mechanics. In our experiment, we directly test the LG inequalities by using a CNOT gate which implements non-invasive measurement under the classical realistic description. This kind of classical non-invasive measurement is also implemented by Knee {\it et al.} \cite{Knee11}. Our method is directly related to the problem of decoherence. By changing the thickness of quartz plates, we can control the evolution time of a single photon between sets of measurements. The ability to violate generalized LG inequality sets the boundary of the classical realistic description.

In our experiment, the coherence length of the initial photon state ($l_{0}$) is about 53 $\mu$m (calculated by $2\pi c/\sigma$). As a result, at the crossover point where the LG inequalities are not violated, the thickness of quartz plate of 33$\lambda_{0}$ corresponds to about 0.486$l_{0}$  (calculated by $(33\lambda_{0}/53)l_{0}$ and $\lambda_{0}$ =0.78$\mu$m). The theoretical form of the output photon state at $t_{2}$ becomes $\rho=0.78|\overline{H}\rangle\langle\overline{H}|+0.22|\overline{V}\rangle\langle\overline{V}|$ with a visibility of 0.56 (the corresponding experimental value is $0.558\pm0.004$). The visibility is calculated by $|p_{|H\rangle}-p_{|V\rangle}|$, where $p_{i}$ represents the corresponding detecting probability ($i\in\{|\overline{H}\rangle,|\overline{V}\rangle\}$). The visibility characterizes the purity of the final state for the measure base is $|\overline{H}\rangle/|\overline{V}\rangle$. When the visibility of the photon state at $t_{2}$ is reduced to less than 0.56, the LG inequalities would not be violated anymore. The state at the transition point where the LG inequalities are not violated still has coherence between the two orthogonal states. It is similar to the case that not all entangled states violate a Bell inequality \cite{Werner89}. Therefore, the ability to violate LG inequalities, which sets the boundary of the classical realistic description, may connect to the ability to perform some quantum information task with quantum advantages as that of Bell inequalities.

In summary, we experimentally violated two generalized LG inequalities in an all-optical system using a CNOT gate. The violation of generalized LG inequalities disproves the definite classical evolution trajectory of the single qubit \cite{Jordan06} and implies that at least one of the two assumptions in the classical realistic description is untenable. The ability to violate LG inequalities can be used to set the boundary of the classical realistic description.
\section{Methods}
In our experiment, the photon of interest was prepared from a heralded single photon source, which was produced from a pulsed parametric down-conversion process. A mode-locked Ti:sapphire laser with a centre wavelength mode locked to 0.78 $\mu$m (130 fs pulse width and 76 MHz repetition rate) was used to pump a 2 mm type-I $\beta$-barium borate (BBO) crystal which generated the second harmonic ultraviolet pulses (0.39 $\mu$m). These ultraviolet pulses were then focused into a 2 mm type-II BBO crystal which was cut for beamlike phase matching \cite{Kurtsiefer01,Takeuchi01} to produced bright down-conversion photon pairs. The evolution of one of the photons was investigated by preparation into $|\overline{H}\rangle$ and passing it through the experimental setup in Fig. 2. The other photon was used as the trigger. We obtained about 18000 coincidence events per second and the integration time was 10 s for each measurement.

\section{Acknowledgements}
We thank Dr. M.-H. Yung for helpful discussion. This work was supported by the National Basic Research Program of China (Grants No. 2011CB9212000), National Natural Science Foundation of China (Grant Nos. 60921091, 10874162, 11004185), China Postdoctoral Science Foundation (Grant No. 20100470836) and the Fundamental Research Funds for the Central Universities (Grant No. WK 2030020019).

\end{document}